\documentclass[twocolumn,showpacs,preprintnumbers,amsmath,amssymb]{revtex4}

\usepackage{graphicx}
\usepackage{dcolumn}
\usepackage{bm}

\begin{document}

\title{Frequency-dependent reflection of spin waves \\from a magnetic inhomogeneity induced by a surface DC-current}

\author{T. Neumann}
 \email{neumannt@physik.uni-kl.de}
\author{A. A. Serga}
\author{B. Hillebrands}
\affiliation{
Fachbereich Physik and Forschungszentrum OPTIMAS\\
Technische Universit\"at Kaiserslautern, 67663 Kaiserslautern, Germany }

\author{M. P. Kostylev}
\altaffiliation{On leave from FET Department, St. Petersburg Electrotechnical University, St.Petersburg
197376, Russia}
\affiliation{
School of Physics\\
University of Western Australia, Crawley, Western Australia 6009, Australia
}%

\date{\today}

\begin{abstract}
The reflectivity of a highly localized magnetic inhomogeneity is experimentally studied. The
inhomogeneity is created by a dc-current carrying wire placed on the surface of a ferrite film. The
reflection of propagating dipole-dominated spin-wave pulses is found to be strongly dependent on the
spin-wave frequency if the current locally increases the magnetic field. In the opposite case  the
frequency dependence is negligible.
\end{abstract}


\maketitle


The interaction of spin-wave packets propagating in a ferromagnetic film with localized artificial
inhomogeneities is of interest for numerous applications. For example, the tunneling \cite{Dem04}, the
guidance \cite{Top08}, the filtering through band gaps \cite{Chu08}, and shaping \cite{Ser07a} of
spin-wave pulses has been realized in this way.

There are different possibilities to create inhomogeneities, such as by mechanical structuring of the
magnetic material \cite{Mae06, Kos08, Chu08} or by variation of the saturation magnetization by ion
irradiation \cite{Car82}. However, these modifications of the magnetic properties are irreversible and
do not allow any adjustment or real-time control.

A more promising approach is to change the magnetic inhomogeneity dynamically as can be done e.g.\ with
parametric pumping by fast variation of the bias magnetic field \cite{Mel99, Ser07a}. In this process
the increase of the precessing transversal magnetization component leads to a decrease of the static
saturation magnetization. However, one should keep in mind the high complexity of the process and the
relatively long time scale in the microsecond range to reach a quasi-equilibrium regime \cite{Dem07,
Neu08}.

Instead, many recent experiments \cite{Kos07, Dem08, Smi08} have relied on the technique to locally
modify the bias magnetic field by the Oersted field of a current-carrying wire placed on the film
surface. With this setup, XOR and NAND gates, milestones in the development of spin-wave logic, have
been realized \cite{Sch08} and resonant spin-wave tunneling has been discovered \cite{Dem07a}. Moreover,
periodic structures, so called magnonic crystals, of such design \cite{Fet04} have the advantage of
being controllable on a time scale shorter than the spin-wave relaxation time.

In the current work, we directly measure the reflected and transmitted intensity of spin-wave pulses
propagating in a thin yttrium-iron-garnet (YIG) film through a current induced magnetic inhomogeneity
for a wide range of spin-wave carrier frequencies and applied dc-currents. The results provide evidence
for the potential of this structure as an effective frequency filter and adjustable energy divider.



\begin{figure}[b]
\includegraphics[height = 38ex]{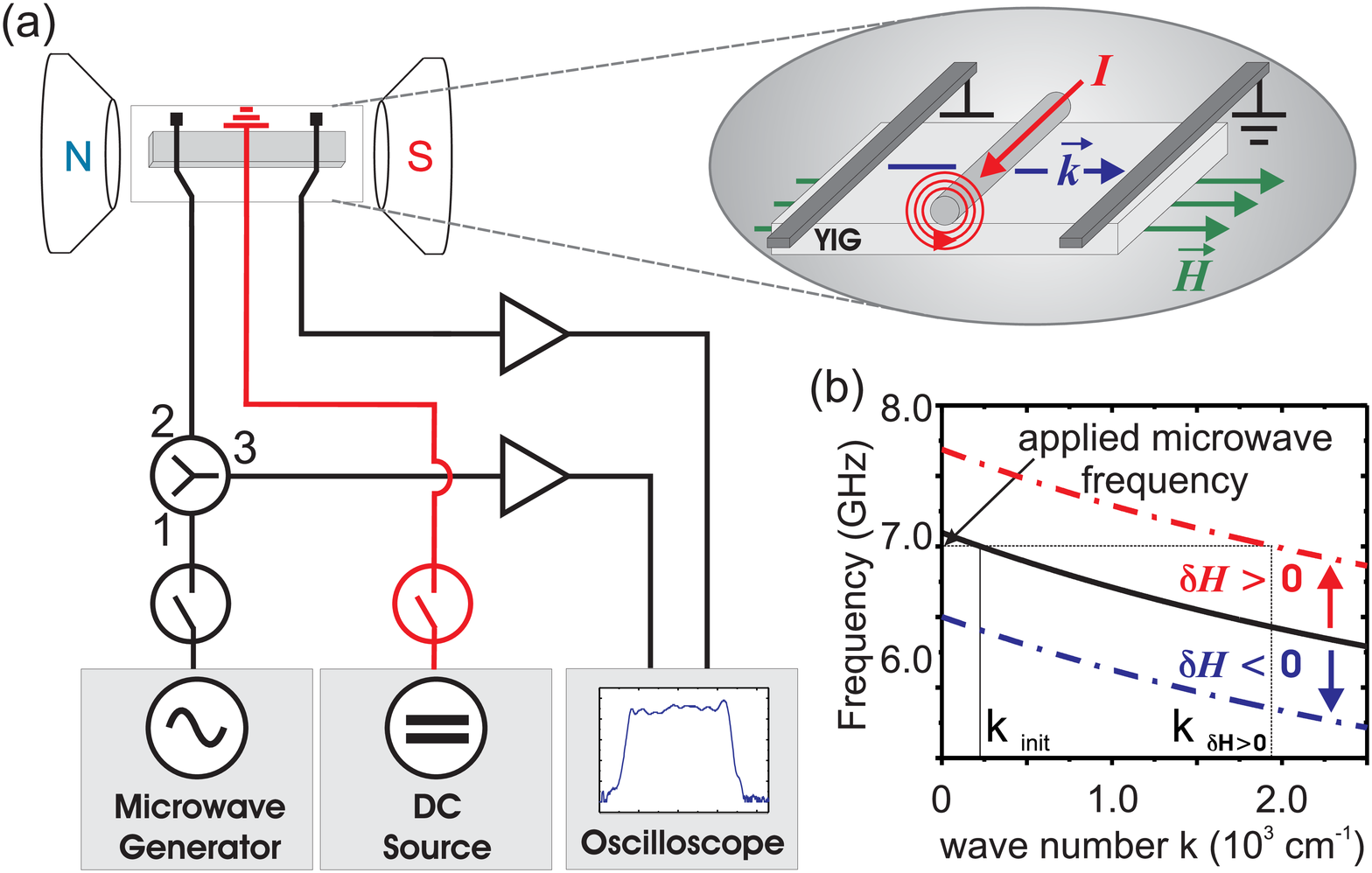}
\caption{\label{fig:aufbau} (Color online) (a) Sketch of the experimental microwave setup and of the
section layout. (b) Schematic representation of the dc-current influence on the dispersion relation.}
\end{figure}

All experiments were performed in the pulse regime. This approach was chosen to check the applicability
of the method for microwave signal processing used in modern digital technique and exclude any influence
of heating effects caused by the dc-current applied to the wire.

Figure~\ref{fig:aufbau}(a) shows a sketch of the experimental setup. A triggered microwave switch
transforms the cw-signal of a microwave generator with a carrier frequency $f$ between $7.010~{\rm GHz}$
and $7.115~{\rm GHz}$ into $280~{\rm ns}$ long pulses with a $1~{\rm ms}$ repetition rate. These pulses
are sent to a $50~{\rm \mu m}$ wide strip-line transducer placed on the surface of a $5~{\rm \mu m}$
thick, longitudinally magnetized single crystal YIG-stripe of $15~{\rm mm}$ width.

The microwave signal excites packets of backward volume magnetostatic waves (BVMSW) whose wave vector is
aligned in the direction of the applied bias magnetic field $H=1800~{\rm Oe}$ with a typical value
between $10~{\rm cm}^{-1}$ and $200~{\rm cm}^{-1}$. The spin waves propagate through the film and are
picked up by a second, identical antenna situated at a distance of $8~{\rm mm}$. By amplifying and
detecting the obtained microwave signal we can observe the transmitted pulse in real time.

While the spin-wave pulse is propagating in the film a $180~{\rm ns}$ long and between $-2~{\rm A}$ and
$+2~{\rm A}$ strong dc-current pulse with a rise time of less than $20~{\rm ns}$ is applied to a
$50~{\rm \mu m}$ thick wire placed on the film surface halfway between the antennae
(Fig.~\ref{fig:aufbau}(a)). It creates a highly localized magnetic inhomogeneity across the
YIG-waveguide of up to $\pm 200~{\rm Oe}$ and width comparable to the spin-wave wavelength. The length
of the dc-current pulse is chosen in order to, on the one hand, avoid heating the sample and, on the
other hand, reach a stationary state of the spin-wave propagation. For such a structure, two principally
different operation regimes exist (Fig.~\ref{fig:aufbau}(b)). If the dc-current locally decreases the
bias magnetic field we operate in the {\it tunneling regime} \cite{Dem04}. We will refer to the opposite
case as {\it diffraction regime}.

Spin waves reflected from the inhomogeneity are picked up by the input antenna. A Y-circulator in the
input channel allows to detect and observe this signal on the oscilloscope simultaneously to the
transmitted one.



\begin{figure}[t]
\includegraphics[height = 74ex]{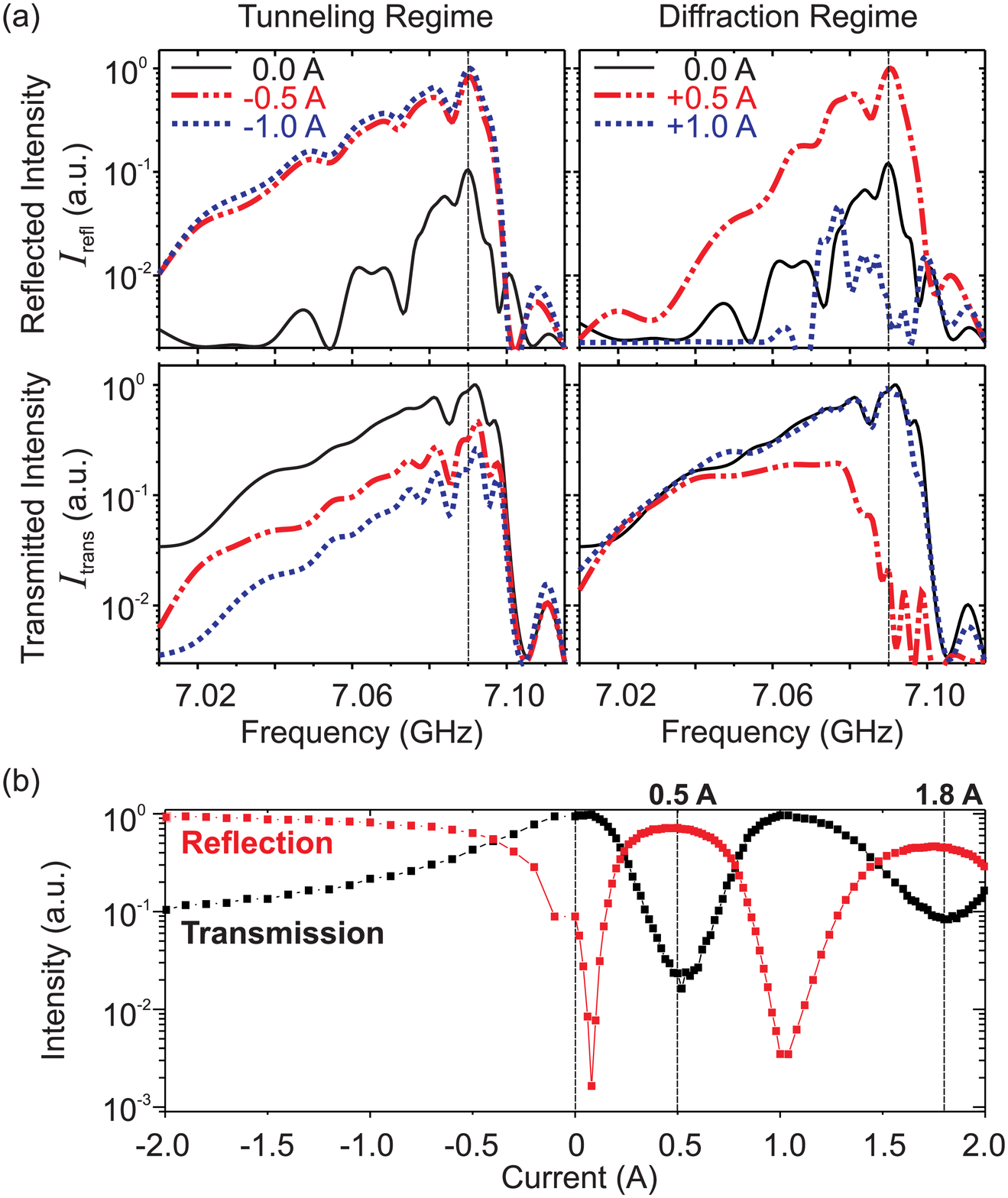}
\caption{\label{fig:results1} (Color online) (a) Reflection and transmission characteristics in the
tunneling and diffraction regime. The vertical dashed line indicates $f=7.09~{\rm GHz}$. (b) Current
dependent reflection and transmission for $f=7.09~{\rm GHz}$.}
\end{figure}
In Fig.~\ref{fig:results1}(a) the transmission and reflection characteristics for exemplary currents
$-0.5~{\rm A}$ and $-1.0~{\rm A}$ representative for the tunneling regime as well as $+0.5~{\rm A}$ and
$+1.0~{\rm A}$ for the diffraction regime are shown together with the characteristics obtained when no
current is applied.

If no current is applied the measured intensities are determined by the excitation and detection
characteristics of the used microstrip antennae and the spin-wave spectrum. Because of the finite size
of the antennae, the excitation and detection is limited to spin waves with a long enough wavelength.
This explains the vanishing intensity for frequencies  below $7.02~{\rm GHz}$. At the same time, the
frequency of ferromagnetic resonance $f_{\rm FMR} \approx 7.10~{\rm GHz}$ 
constitutes an upper frequency limit for the BVMSW in the dipole approximation. Hence, the intensity of
the detected signal quickly drops above it. Note, that in the absence of a current, only a relatively
small part of the spin wave is reflected by the metal wire placed on the sample surface \cite{Mel87}, so
that most of the spin-wave signal is observed in transmission. The dips observable in the reflected and
transmitted signal are caused by resonances between the input antenna and the central wire.

When a dc-current is applied, the ratio between reflection and transmission changes significantly
depending on the direction and magnitude of the applied dc-current.
In the tunneling regime, i.e. when the bias magnetic field is locally decreased by the current, the
reflected signal monotonically increases over the whole investigated frequency range with increasing
current modulus.
In the diffraction regime the behavior is non-monotonous. When the current is increased two reflection
resonances are clearly observed. They are most pronounced for $f = 7.09~{\rm GHz}$, which is slightly
below $f_{\rm FMR}$ (see Fig.~\ref{fig:results1}).

For $0.5~{\rm A}$ the reflected signal intensity increases drastically compared to the case without
current. The reflected signal rises within $30~{\rm ns}$ after the application of the dc-pulse by a
factor of $10$. The intensity then stays constant for the remainder of the microwave signal pulse.

Further increase of the current leads to a decrease of the reflected signal intensity. At a current of
$1.0~{\rm A}$ the reflected signal is effectively suppressed and the transmitted pulse shape is almost
undisturbed. This behavior is repeated when the current is further increased. The reflected spin-wave
intensity rises till it reaches a second maximum at $1.8~{\rm A}$ and then drops again.
Fig.~\ref{fig:results1}(b) displays the current dependence of the reflected and transmitted signal
intensity for $f = 7.09~{\rm GHz}$.

By adding the reflected and transmitted signal intensities, it can be verified that their sum stays
constant.

To explain the two different regimes, consider Fig.~\ref{fig:aufbau}(b) again. In the tunneling regime,
for large enough current modulus a barrier is formed which reflects the signal. A larger absolute
current leads to an increase of the zone where the spin-wave propagation is prohibited and through
which, consequently, the spin waves cannot propagate and need to tunnel \cite{Dem04}. In the diffraction
regime, constructive interference of spin waves reflected from the region of inhomogeneous magnetic
field occurs. The observed data on spin-wave reflection supports very well the theoretical prediction
reported in \cite{Kos07} by the authors.

\begin{figure}[t]
\includegraphics[height = 28ex]{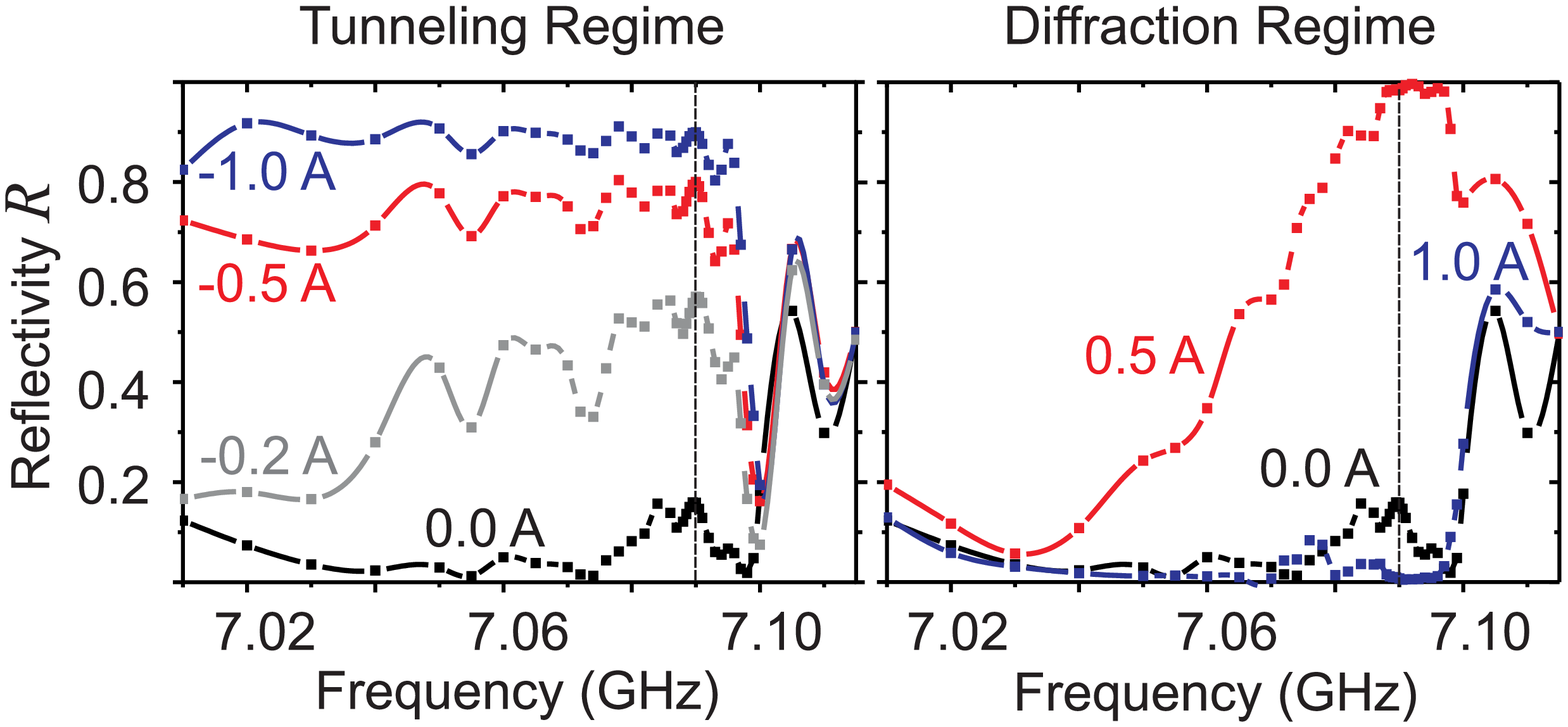}
\caption{\label{fig:reflectivity} (Color online) Frequency-dependent reflectivity.}
\end{figure}

We determine the frequency-dependent reflectivity $R$ of the investigated structure via
\begin{displaymath}
R = \frac{I_{\rm refl}}{I_{\rm trans}+I_{\rm refl}}
\end{displaymath}
where $I_{\rm refl}$ and $I_{\rm refl}$ denote the reflected respectively transmitted intensity
(Fig.~\ref{fig:reflectivity}). In the tunneling regime, for large enough currents the reflectivity is
constant over the whole range of accessible frequencies. The picture is completely different in the
diffraction regime. Here, the reflectivity is clearly frequency-dependent. For low frequencies between
$7.01~{\rm GHz}$ and $7.04~{\rm GHz}$ it is constant and restricted to $R<0.2$. For higher frequencies
which are closer to $f_{\rm FMR}$, a strong dc-current dependence is observed. In particular, we note
that the reflectivity for $f=7.09~{\rm GHz}$ and $I=-0.5~{\rm A}$ is only $0.78$ while it reaches almost
$1$ for the opposite dc-current polarity.


In conclusion, we have investigated the reflection of dipole-dominated spin waves propagating through a
highly localized dc-current induced magnetic inhomogeneity. The reflection on the inhomogeneity created
by a surface dc-current proves to be dependent on the frequency of the propagating spin wave as well as
on the magnitude and polarity of the dc-current. While the latter two dependencies are monotonous for a
current decreasing the local magnetic field ({\it tunneling regime}) they exhibit a resonant structure
in the opposite case ({\it diffraction regime}) which is well pronounced only in a small frequency range
below $f_{\rm FMR}$. For a given current of $0.5~{\rm A}$ the ratio of transmitted signals in tunneling
and diffraction operation regime is more than $25$.

These results have to be considered for the future design of current controlled spin-wave devices, e.g.\
spin-wave logic gates and dynamic magnonic crystals. While the diffraction regime has the advantage of a
high reflectivity for relatively low and easily reachable currents (which allow for long current pulses
without disturbing heating effects), the tunneling regime allows the design of frequency-independent
structure. This may be especially interesting for ultra-short pulses with a wide Fourier spectrum. In
addition, the frequency-dependent reflection in the diffraction regime can be used to create tunable
frequency selective devices. By using multiple wires instead of just a single one, the characteristics
of the structure can be further improved.

Financial support by the MATCOR Graduate School of Excellence, the DFG SE 1771/1-1, the Australian
Research Council, and the University of Western Australia is gratefully acknowledged.


\begin{thebibliography}{1}

\bibitem{Dem04} S.~O. Demokritov, A.~A. Serga, A. Andr\'e, V.~E. Demidov, M.~P. Kostylev, and B. Hillebrands, Phys. Rev. Lett. {\bf 93}, 047201 (2004).

\bibitem{Top08} J. Topp, J. Podbielski, D. Heitmann, and D. Grundler, Phys. Rev. B {\bf 78}, 024431 (2008).

\bibitem{Chu08} A.~V. Chumak, A.~A.Serga, B. Hillebrands, and M.~P. Kostylev, Appl. Phys. Lett. {\bf 93}, 022508 (2008).

\bibitem{Ser07a} A.~A. Serga, T. Schneider, B. Hillebrands, M.~P. Kostylev, and A.~N. Slavin, Appl. Phys. Lett. {\bf 90}, 022502 (2007).

\bibitem{Mae06} A. Maeda, and M. Susaki, IEEE Trans. Magn. {\bf 42}, 3096 (2006).

\bibitem{Kos08} M.~P. Kostylev, P. Schrader, R.~L. Stamps, G. Gubbiotti, G. Carlotti, A.~O. Adeyeye, S. Goolaup, and N. Singh, Appl. Phys. Lett. {\bf 92}, 132504 (2008).

\bibitem{Car82} R.~L. Carter, J.~M. Owens, C.V. Smith, and K.W. Reed, J. Appl. Phys. {\bf 53}, 2655 (1982).

\bibitem{Mel99} G.~A. Melkov, A.~A. Serga, A.~N. Slavin, V.~S. Tiberkevich, A.~N. Oleinik, and A.~V. Bagada, JETP {\bf 89}, 1189 (1999).

\bibitem{Dem07} V.~E. Demidov, O. Dzyapko, S.~O. Demokritov, G.~A.  Melkov, and A.~N. Slavin, Phys. Rev. Lett. {\bf 99}, 037205 (2007).

\bibitem{Neu08} T. Neumann, A.~A. Serga, and B. Hillebrands, submitted, arXiv:0810.4033v1 [cond-mat.other] (preprint).

\bibitem{Kos07} M.~P. Kostylev, A.~A. Serga, T. Schneider, T. Neumann, B. Leven, B. Hillebrands, and R.~L. Stamps, Phys. Rev. B {\bf 76}, 184419 (2007).

\bibitem{Dem08} V.~E. Demidov, U.~H. Hansen, and S.~O. Demokritov, Phys. Rev. B {\bf 78}, 054410 (2008).

\bibitem{Smi08} K.~R. Smith, M.~J. Kabatek, P. Krivosik, and M. Wu, J. Appl. Phys.  {\bf 104}, 043911 (2008).

\bibitem{Sch08} T. Schneider, A.~A. Serga, B. Leven, B. Hillebrands, R.~L. Stamps, and M.~P.Kostylev, Appl. Phys. Lett. {\bf 92}, 022505 (2008).

\bibitem{Dem07a} U. Hansen, M. Gatzen, V.~E. Demidov, and S.~O. Demokritov, Phys. Rev. Lett. {\bf 99}, 127204 (2007).

\bibitem{Fet04} Y.~K. Fetisov, J. Comm. Tech. Electron. {\bf 10}, 1171 (2004).

\bibitem{Mel87} I.~V. Krutsenko, G.~A. Melkov, and S.~A. Ukhanov, Radiotexn. i Elektron. {\bf 32}, 1976 (1987).

\end{thebibliography}
\end{document}